\documentclass[apj]{emulateapj}
\submitted{Submitted to The Astrophysical Journal Letters on 2016, November 11; accepted for publication on 2016, December 7. }

\usepackage{color}

\newcommand\ionB[2]{#1$\;${\scshape{#2}}}

\def\FFeII{\mbox{[\ionB{Fe}{ii}]}}

\def\Ha{\ifmmode^{\mathrm{H}\alpha }\else$\mathrm{H}\alpha$\fi}
\def\Hb{\ifmmode^{\mathrm{H}\beta }\else$\mathrm{H}\beta$\fi}
\def\LyA{\ifmmode^{\mathrm{H}\alpha }\else$\mathrm{Ly}\alpha$\fi}
\def\BrA{\ifmmode^{\mathrm{Br}\alpha }\else$\mathrm{Br}\alpha$\fi}
\def\BrG{\ifmmode^{\mathrm{Br}\gamma }\else$\mathrm{Br}\gamma$\fi}
\def\PaB{\ifmmode^{\mathrm{Pa}\beta }\else$\mathrm{Pa}\beta$\fi}
\def\mag{\ifmmode^{\rm m }\else$^{\rm m}$\fi}

\def\as{$\,^{\prime\prime}\,$}

\def\hh{\ifmmode^{\rm h}\else$^{\rm h}$\fi}
\def\mm{\ifmmode^{\rm m}\else$^{\rm m}$\fi}
\def\ss{\ifmmode^{\rm s}\else$^{\rm s}$\fi}
\def\deg{\ifmmode^\circ\else$^\circ $\fi}
\def\amin{\ifmmode^\prime\else$^\prime $\fi}

\def\decdm#1#2{\ifmmode{#1}\else{$#1$}\fi\deg\ #2\amin\ }

\def\dec#1#2#3{\ifmmode{#1}\else{$#1$}\fi\deg\ #2\amin\ #3\as\ }

\def\decb#1#2#3#4{\ifmmode{#1}\else{$#1$}\fi\deg\ #2\amin\ #3\farcs#4 }

\shorttitle{A high-mass protobinary system with spatially resolved circumstellar disks \& circumbinary disk}
\shortauthors{Kraus et al.}

\begin{document}


\title{A high-mass protobinary system with spatially resolved circumstellar accretion disks and circumbinary disk}

\footnotetext[$\star$]{Based on observations made with ESO telescopes 
at Paranal Observatory under program IDs 
60.A-9174(A),
089.C-0819(A,C),
089.C-0959(D,E),
094.C-0153(A),
096.C-0652(A).
}


\author{
S.~Kraus\altaffilmark{1}, 
J.~Kluska\altaffilmark{1},
A.~Kreplin\altaffilmark{1},
M.~Bate\altaffilmark{1},  
T.\ J.~Harries\altaffilmark{1},
K.-H.~Hofmann\altaffilmark{2},
E.~Hone\altaffilmark{1},
J.\ D.~Monnier\altaffilmark{3},
G.~Weigelt\altaffilmark{2},
N.~Anugu\altaffilmark{1},
W.J.~de~Wit\altaffilmark{4},
M.~Wittkowski\altaffilmark{5}
}

\email{skraus@astro.ex.ac.uk}

\affil{
$^{1}$~School of Physics, Astrophysics Group, University of Exeter, Stocker Road, Exeter EX4 4QL, UK\\
$^{2}$~Max-Planck-Institut f\"ur Radioastronomie, Auf dem H\"ugel 69, 53121 Bonn, Germany\\
$^{3}$~Department of Astronomy, University of Michigan, 311 West Hall, 1085 South University Ave, Ann Arbor, MI 48109, USA\\
$^{4}$~ESO, Alonso de Cordova 3107, Vitacura, Santiago 19, Chile\\
$^{5}$~ESO, Karl-Schwarzschild-Str.\ 2, 85748 Garching bei M\"unchen, Germany
}


\begin{abstract}
  High-mass multiples might form via fragmentation 
  of self-gravitational disks or alternative scenarios such as disk-assisted capture.
  However, only few observational constraints exist on the architecture and disk structure of 
  high-mass protobinaries and their accretion properties. 
  Here we report the discovery of a close ($57.9\pm0.2$mas=170au) 
  high-mass protobinary, IRAS17216-3801, where our 
  VLTI/GRAVITY+AMBER near-infrared interferometry
  allows us to image the circumstellar disks around the individual components
  with $\sim3$\,milliarcsecond resolution.
  We estimate the component masses to $\sim20$ and $\sim18M_{\sun}$
  and find that the radial intensity profiles can be reproduced
  with an irradiated disk model, where the inner regions are
  excavated of dust, likely tracing the dust sublimation region in these disks.
  The circumstellar disks are strongly misaligned with respect to the binary separation vector, 
  which indicates that the tidal forces did not have time to realign the disks,
  pointing towards a young dynamical age of the system. 
  We constrain the distribution of the Br$\gamma$ and CO-emitting gas
  using VLTI/GRAVITY spectro-interferometry and VLT/CRIRES spectro-astrometry 
  and find that the secondary is accreting at a higher rate than the primary.
  VLT/NACO imaging shows $L'$-band emission on 
  $3-4\times$ larger scales than the binary separation,
  matching the expected dynamical truncation radius for the circumbinary disk.
  The IRAS17216-3801 system is $\sim3\times$ more massive and $\sim5\times$ more compact than
  other high-mass multiplies imaged at infrared wavelengths
  and the first high-mass protobinary system where circumstellar and circumbinary 
  dust disks could be spatially resolved.
  This opens exciting new opportunities for studying 
  star-disk interactions and the role of multiplicity in high-mass star formation.
\end{abstract}


\section{Introduction}

There is now solid evidence that the formation of high-mass stars ($>10M_{\sun}$)
involves accretion through circumstellar disks \citep{ces05,kra10,bol13,joh15}.
This observational evidence supports recent theoretical work which suggests
that the radiation pressure barrier problem can be overcome when considering 
more complex than spherically symmetric infall geometries \citep{yor02,kru09}.
However, the classical accretion disk scenario, which assumes the monolithic collapse
of a core to a single star and which is commonly applied to low-mass star formation, 
fails to explain the high multiplicity fraction that has been observed for high-mass stars.
Multiplicity studies on main-sequence stars found that $\gtrsim80$\%
of all O-type stars ($M\gtrsim16M_{\sun}$) are close multiple systems, 
while this fraction rapidly drops to 20\% for stellar masses of $\sim3M_{\sun}$ \citep{chi12}.
At the same time, the number of companions per system also increases with
stellar mass.  For instance, the five O-/early B-type stars in the 
Orion Trapezium cluster have on average 2.5 known companions, which is about
$5\times$ higher than what has been found for low-mass stars \citep{gre13}.
Different mechanisms have been proposed in order to explain 
these extraordinary characteristics. 
For instance, these high-mass multiples might form via fragmentation of 
self-gravitating massive disks \citep{kra06},
disk-assisted capture \citep{cla91,bal05}, or through failed mergers in stellar collisions \citep{dal06}.
Testing these scenarios requires the detection of high-mass ``protobinaries''
that are still in their formation phase.

Here we report high-angular resolution observations that resolve
the high-mass protostar IRAS17216-3801 into a close ($\sim170$au) binary,
where both components are associated with actively accreting circumstellar disks.
IRAS17216-3801 (=G350.011-01.341) was identied as a high-mass YSO by \citet{per87}
and its distance estimated to $3.08\pm0.6$\,kpc \citep{bol13}.
Various studies reported the detection of OH masers \citep{coh88,arg00,cas98}. 
\citet{bol13} resolved the mid-infrared (8-13$\mu$m) emitting-region
and found an elongated structure with a Gaussian full-width-at-half-maximum (FWHM)
size of $\sim85^{+20}_{-33}$\,milliarcseconds (mas) and speculated that the 
emitting structure might be associated with a circumstellar disk or an outflow.

We observed the source as part of a small survey on high-mass YSOs that we conducted 
with the Very Large Telescope Interferometer (VLTI) 
and its AMBER beam combiner. 
The AMBER measurements with low spectral resolution ($R=\lambda/\Delta\lambda=35$) 
were complemented with higher spectral resolution data ($R=500$)
from the newly-commissioned VLTI/GRAVITY combiner.

\begin{figure*}[p]
  \centering
  $\begin{array}{r@{\hspace{5mm}}c}
    \includegraphics[angle=0,scale=0.83]{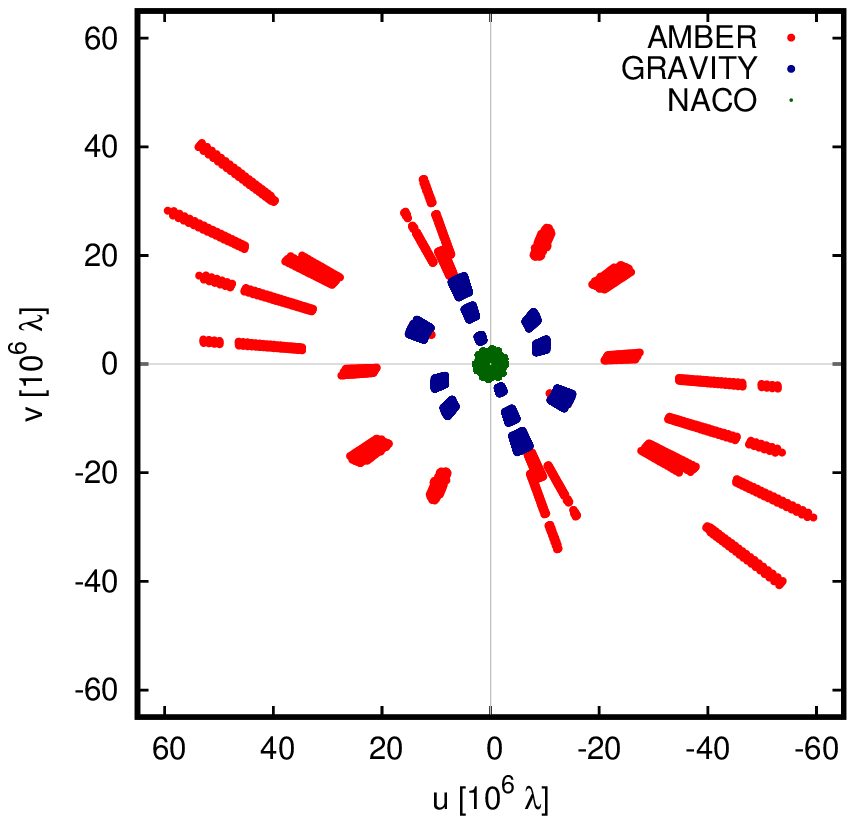} &
    \includegraphics[angle=0,scale=0.71]{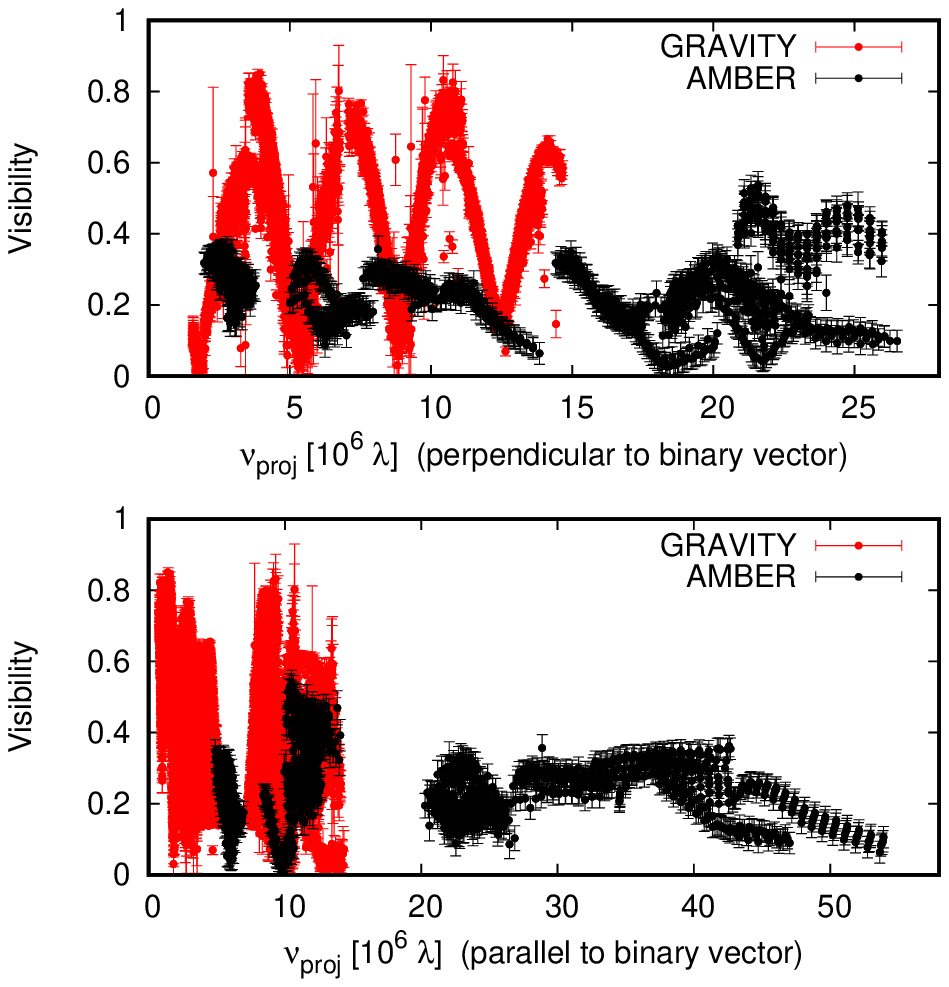} \\[5mm]
    \includegraphics[angle=0,scale=0.55,trim=0cm -1.3cm 3cm 0cm,clip=true]{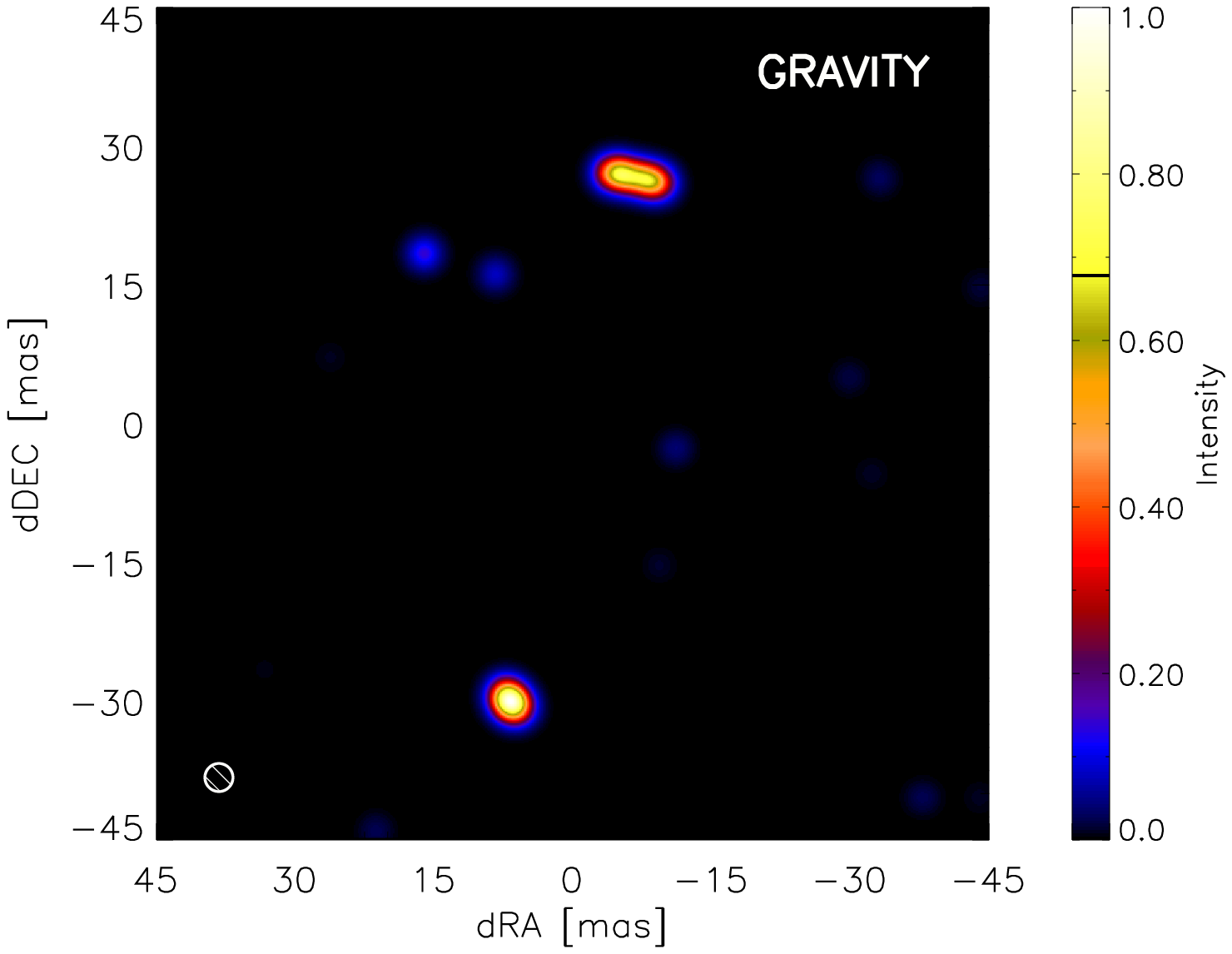} &
    \includegraphics[angle=0,scale=0.55,trim=0cm -1.3cm 0cm 0cm,clip=true]{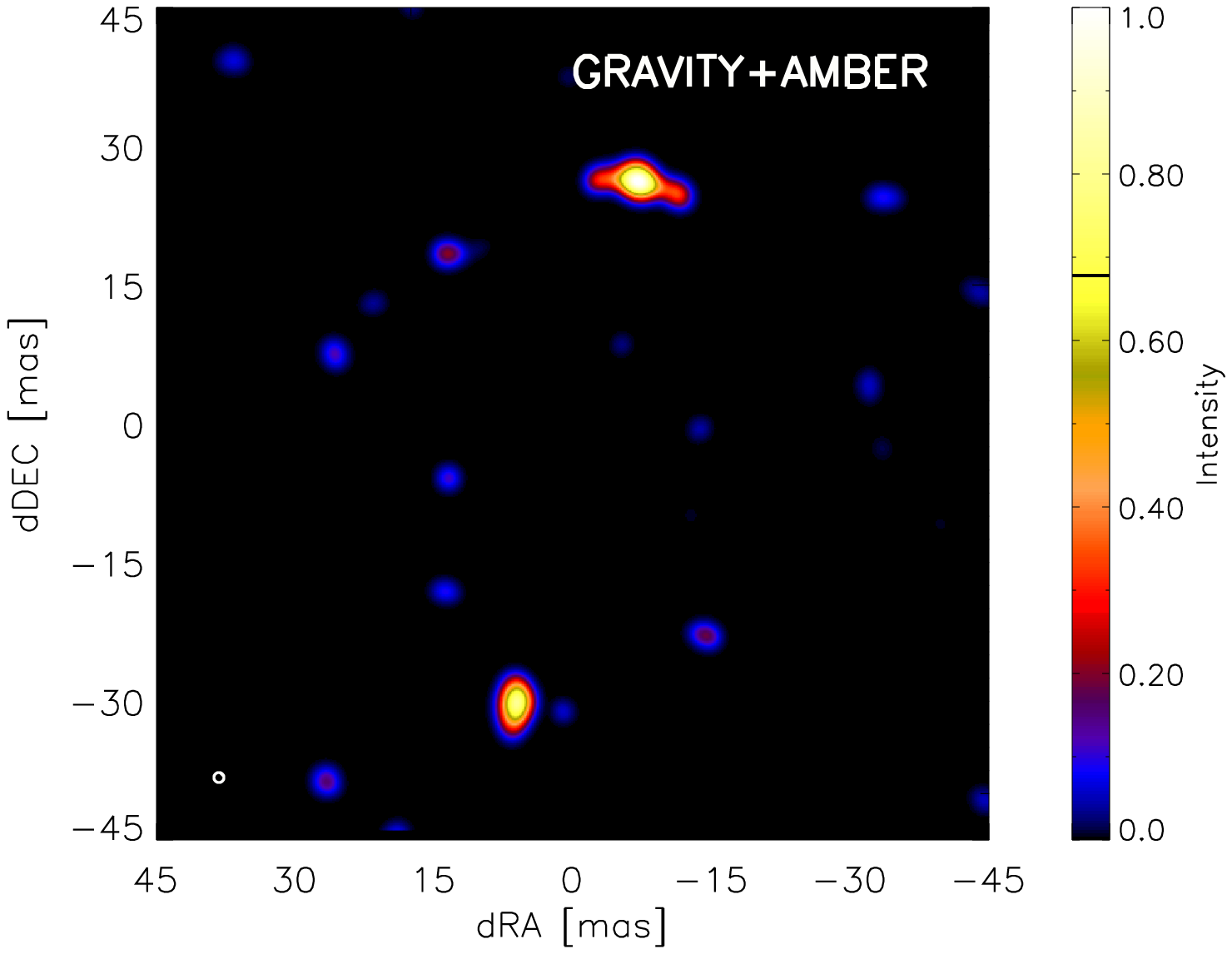}\\
  \end{array}$
  \caption{ 
    Spectro-interferometry on IRAS17216-3801: 
    Obtained $uv$-coverage (top-left) and the measured visibility amplitudes, 
    plotted as function of spatial frequency projected perpendicular (top-right, upper panel) and parallel to the binary separation vector (top-right, lower panel).
    Bottom: Aperture synthesis images obtained with GRAVITY (bottom-left) and GRAVITY+AMBER (bottom-right).\label{fig:VLTI}}
\end{figure*}

\begin{figure*}[p]
  \centering 
  $\begin{array}{l@{\hspace{5mm}}c}
     \includegraphics[angle=0,scale=0.5]{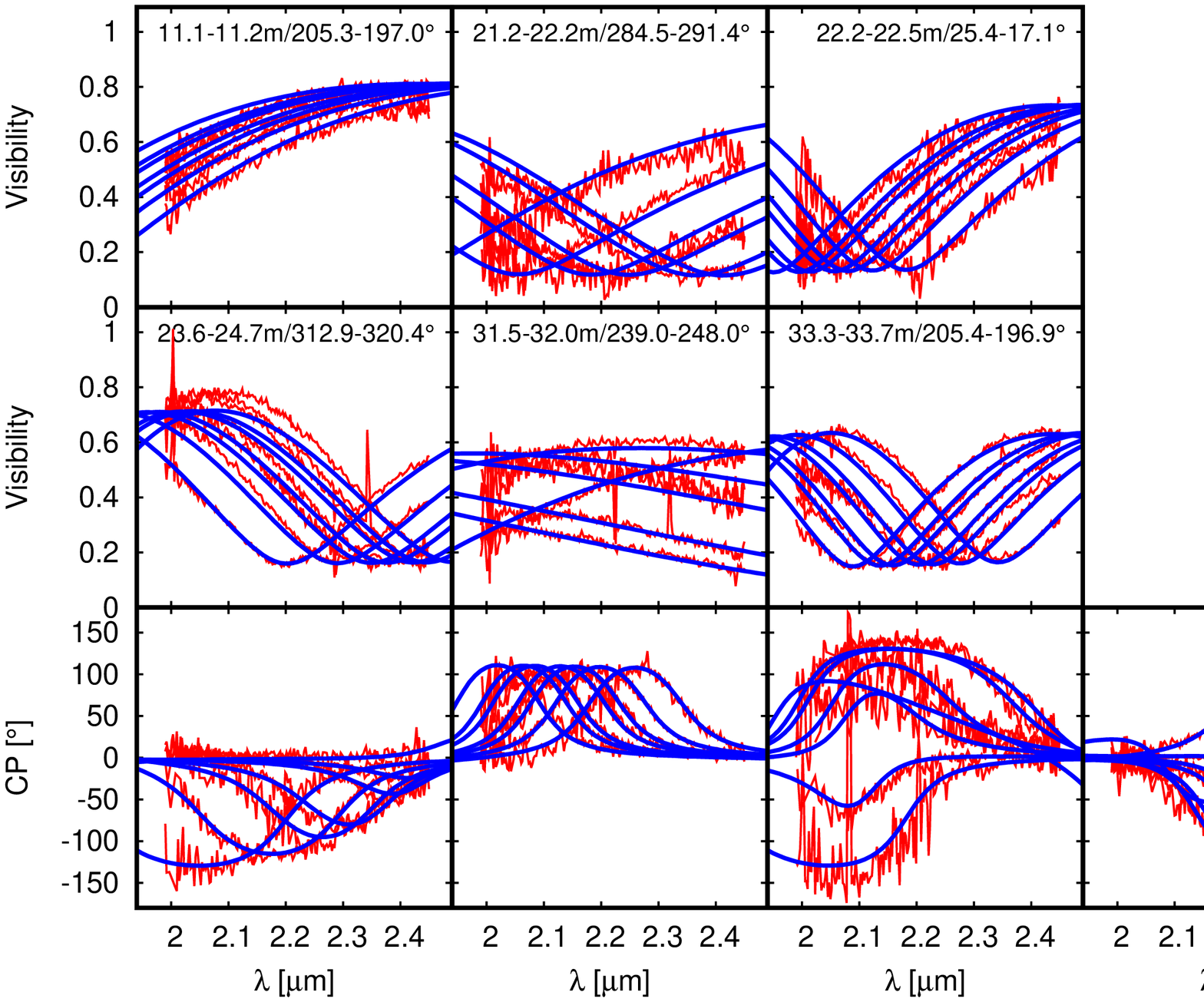} &  \includegraphics[angle=0,scale=0.2]{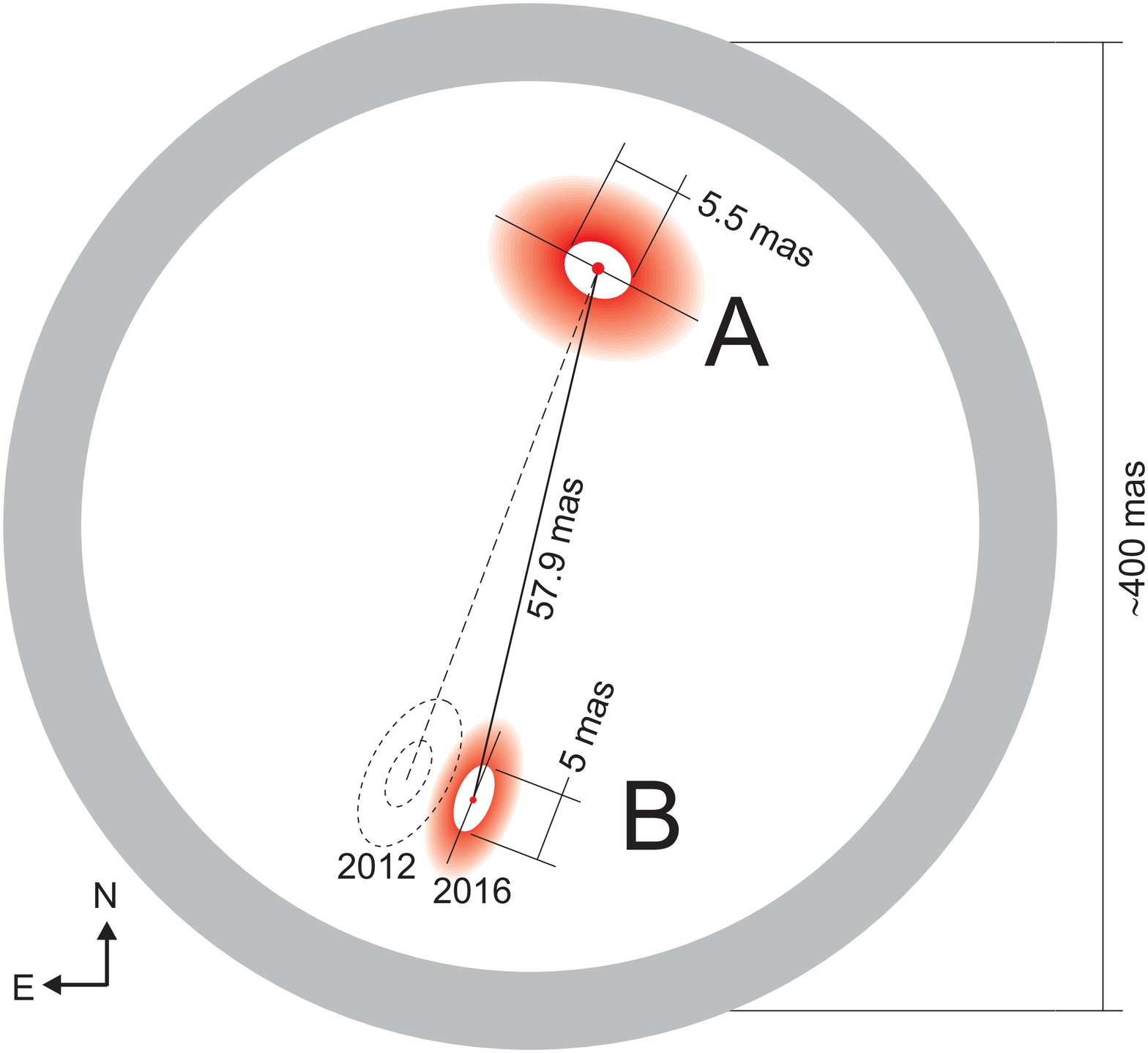} \\[5mm]
     \includegraphics[angle=0,scale=0.5]{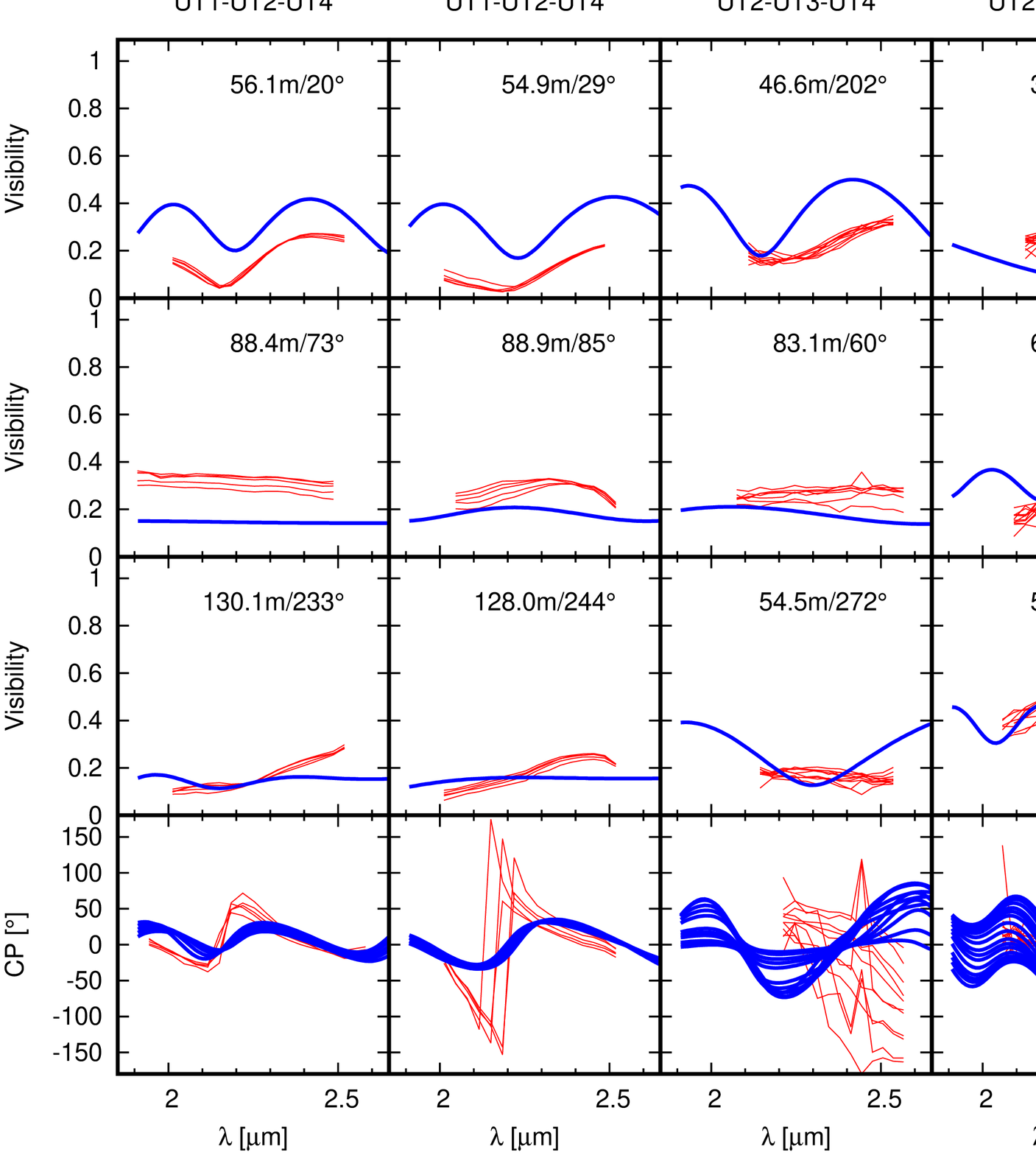} &
    \\
  \end{array}$
  \caption{ 
    GRAVITY (top panel; red lines, 300s bins) and AMBER data (bottom panel; red lines, 50s bins), 
    overplotted with the best-fit model (blue lines).
    In the top-right corner, we illustrate the model components (sketch not to scale).\label{fig:data}}
\end{figure*}

\section{Observations}
\label{sec:observations}

\subsection{VLTI/AMBER spectro-interferometry}

We observed IRAS17216-3801 on 
2012-05-06, 2012-06-01, and 2012-06-03 using the AMBER instrument \citep{pet07}.
The observations used the UT1-UT2-UT4 and UT2-UT3-UT4 triplets of the 
VLTI array of four 8.2m unit telescopes, with projected 
baseline lengths 30...130m (Fig.~\ref{fig:VLTI}, top-left).

Our setup covered the $K$-band with $R=35$, where we 
employed a 50ms integration time.  
The science star observations were bracketed with observations on 
calibrator stars with known diameters, which allowed us to correct 
for atmospheric and instrumental effects.
As calibrators we selected HD155259 (2012-05-06, 2012-06-01, UDD=$0.226\pm0.016$mas)
and HD165787 (2012-05-06, 2012-06-03, UDD=$0.49\pm0.04$mas).
The uniform disk diameters (UDD) were estimated using the JMMC SearchCal tool \citep{bon11}.

The visibilities and closure phases were extracted using the
amdlib software \citep[Release 3.0.4;][]{tat07b,che09}.
The resulting amplitudes/phases show 
strong wavelength-differential modulations (Fig.~\ref{fig:data}, bottom), 
which is characteristic for wide-separation multiple systems. 
Furthermore, we noticed that the observables, in particular
the closure phases, changed slightly between individual object exposures.
Therefore, we decided to analyse the individual exposures ($\sim50$s)
separately in order to avoid temporal smearing. 
We also checked whether our data might be affected by 
bandwidth smearing effects using the procedure outlined in \citet{kra05}.
We estimate the loss of coherence to $\lesssim2.5$\% for our longest baseline, 
which is within our measurement uncertainty.

\subsection{VLTI/GRAVITY spectro-interferometry}

On 2016-06-23 we acquired observations with the
VLTI/GRAVITY instrument \citep[][Eisenhauer et al., in prep.]{eis11} 
as part of Science Verification.
The observations combined the light from the four 
VLTI 1.8m auxiliary telescopes and covered the $K$-band 
with $R=500$.
The telescopes were in the compact configuration (A0-B2-D0-C1)
with projected baseline lengths 11...34m.
Seven object exposures were recorded on IRAS17216-3801,
where each exposure consists of 
30~interferograms taken with 10s integration time.

The science star observations were bracketed with 
observations on calibrator HD159868 (UDD=$0.358\pm0.025$mas).
Wavelength-differential visibilities and phases were extracted using the 
GRAVITY pipeline \citep[Release 0.8.4;][]{lap14}.
Over the 66\,minutes covered by our data recording sequence,
we see strong visibility and phase changes.  
Therefore, we splitted the data into bins and reduced them separately,
where we use 50s binning to minimize 
potential temporal smearing effects in our model fits.

\subsection{VLT/NACO imaging}

\begin{figure*}[p]
  \centering
  $\begin{array}{c@{\hspace{1mm}}c@{\hspace{1mm}}c@{\hspace{1mm}}c}
    \includegraphics[angle=0,scale=0.33, trim=-1.1cm -1.3cm 3.1cm 0cm,clip=true]{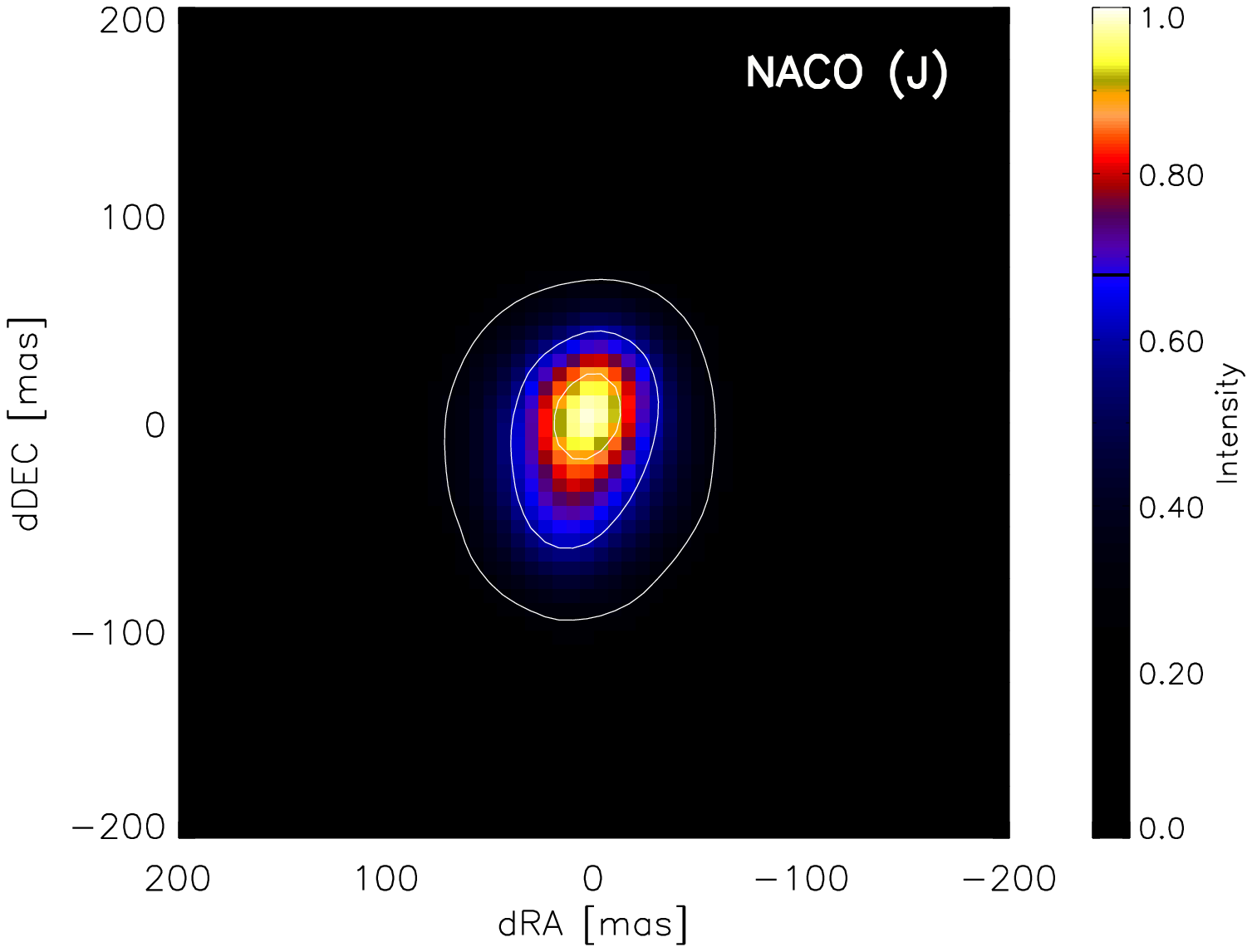} &
    \includegraphics[angle=0,scale=0.33, trim=0.4cm -1.3cm 3.1cm 0cm,clip=true]{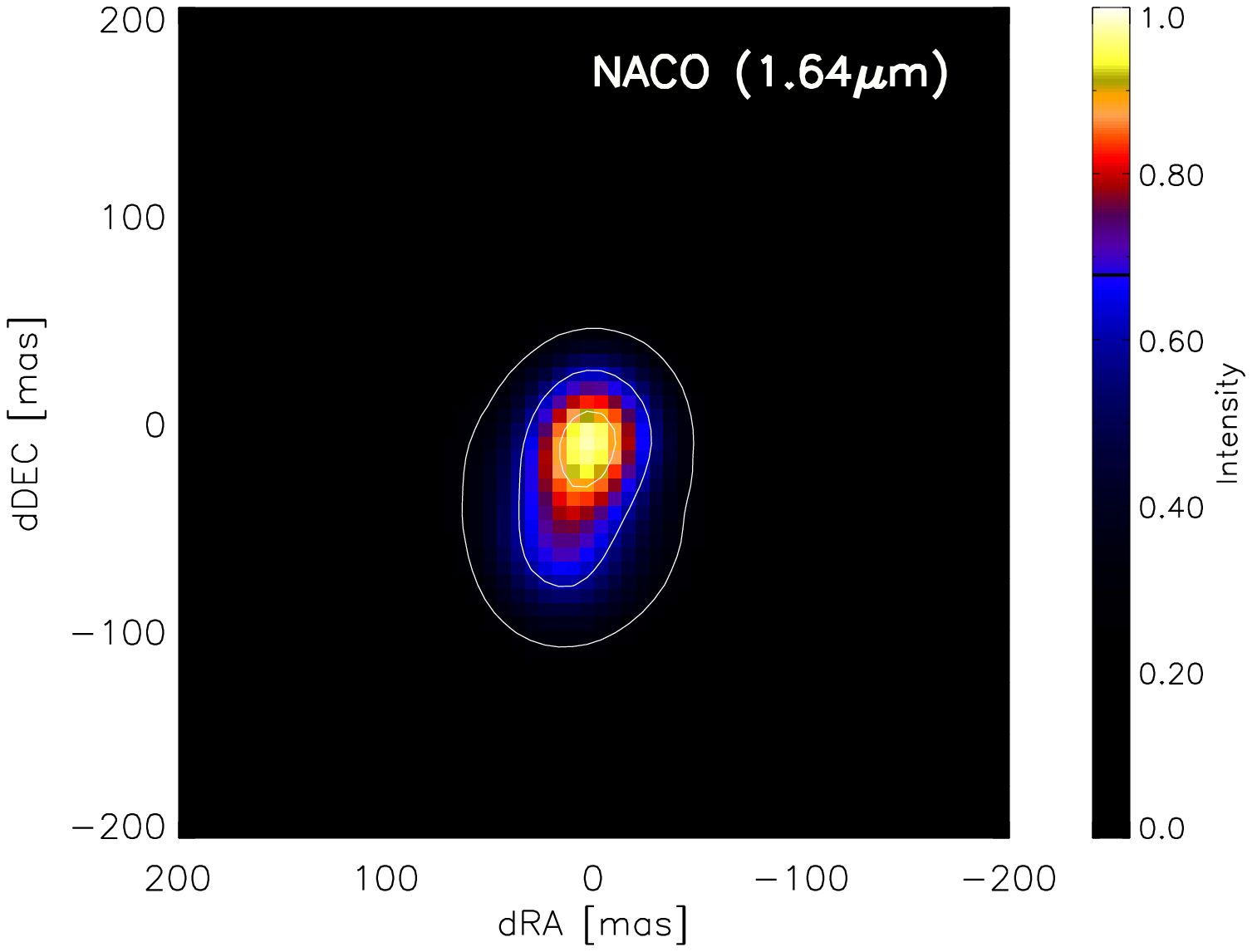} &
    \includegraphics[angle=0,scale=0.33, trim=0.4cm -1.3cm 3.1cm 0cm,clip=true]{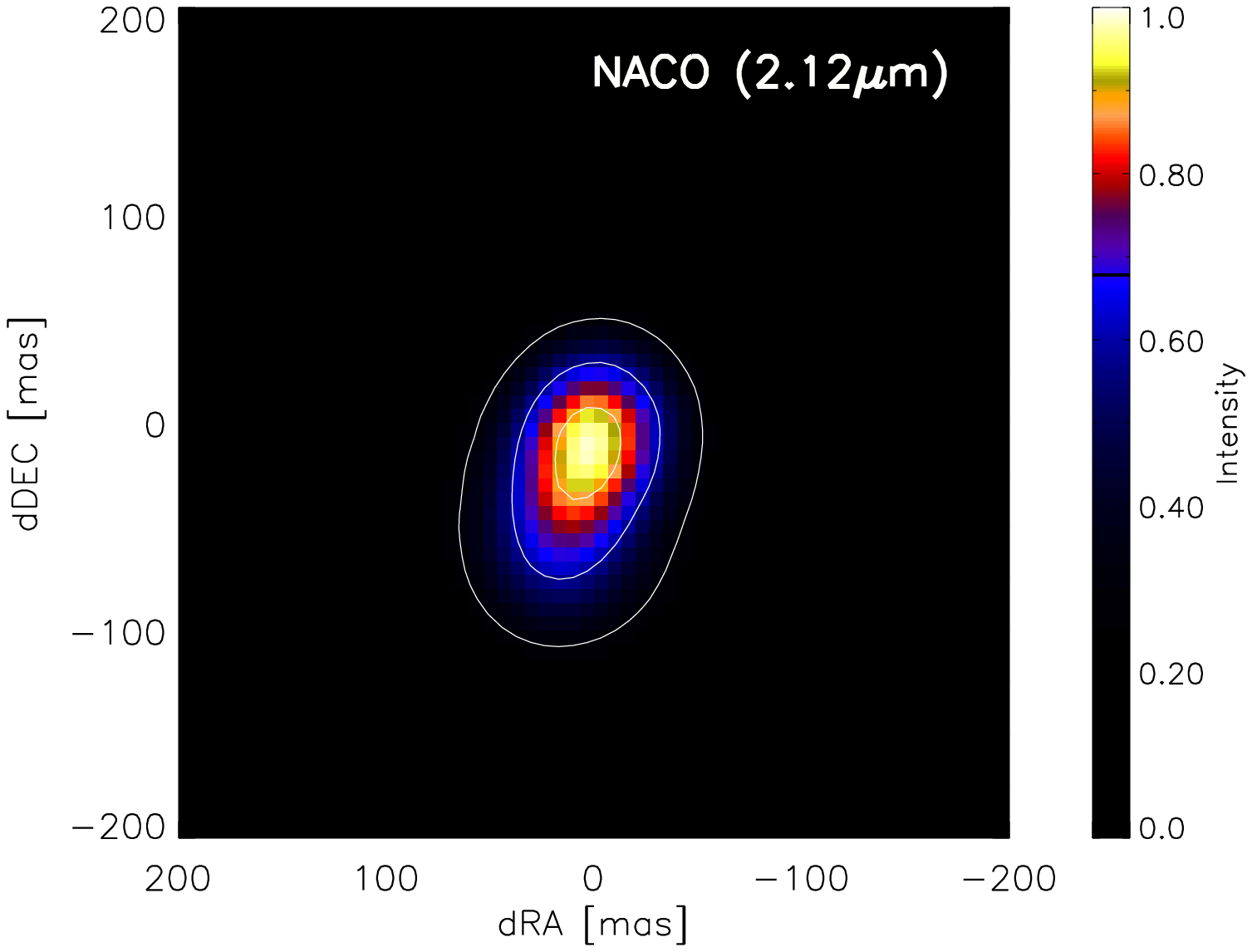} &
    \includegraphics[angle=0,scale=0.33, trim=0.4cm -1.3cm 0cm 0cm,clip=true]{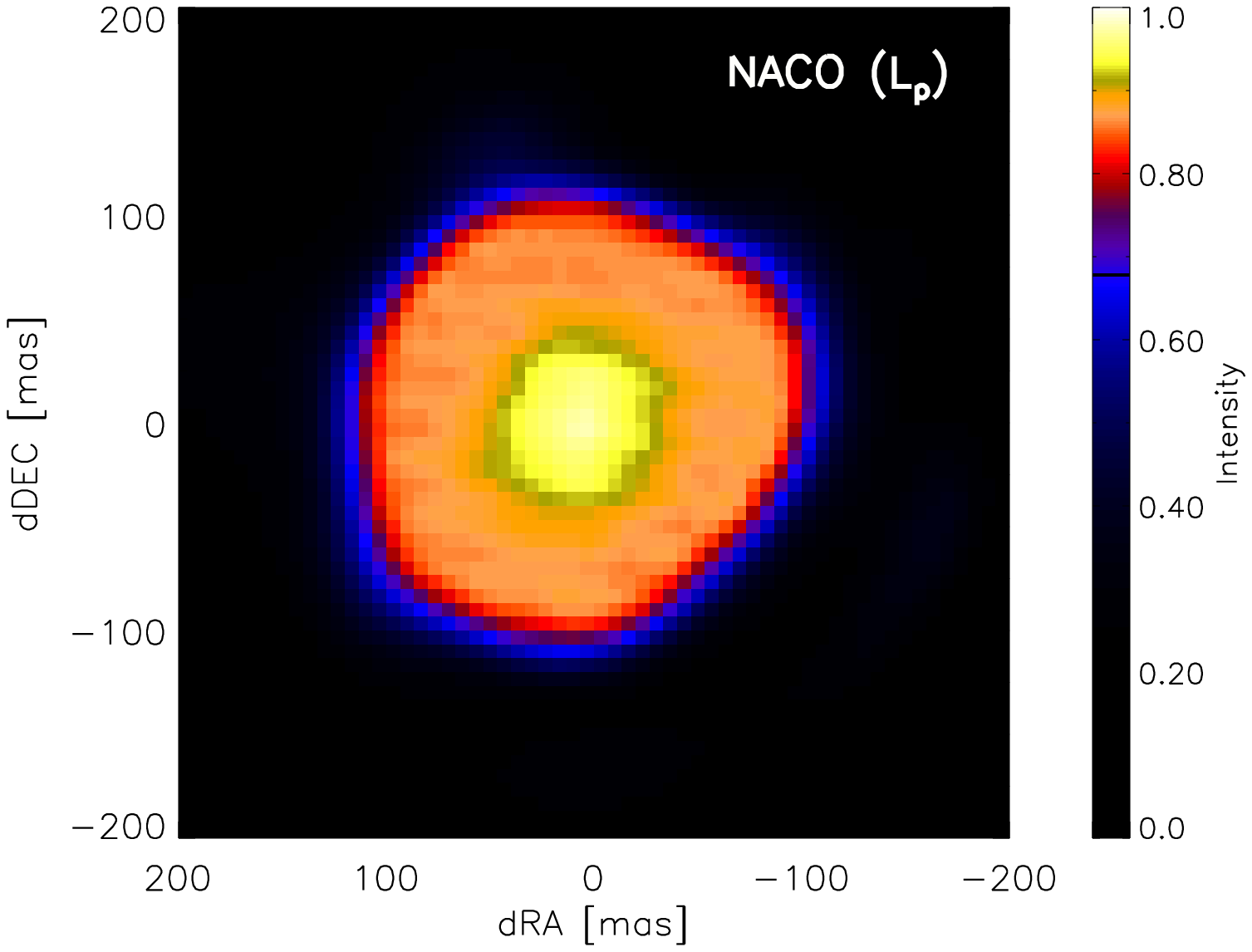} \\[5mm]
    \includegraphics[angle=0,scale=0.33, trim=0.4cm -1.3cm 3.1cm 0cm,clip=true]{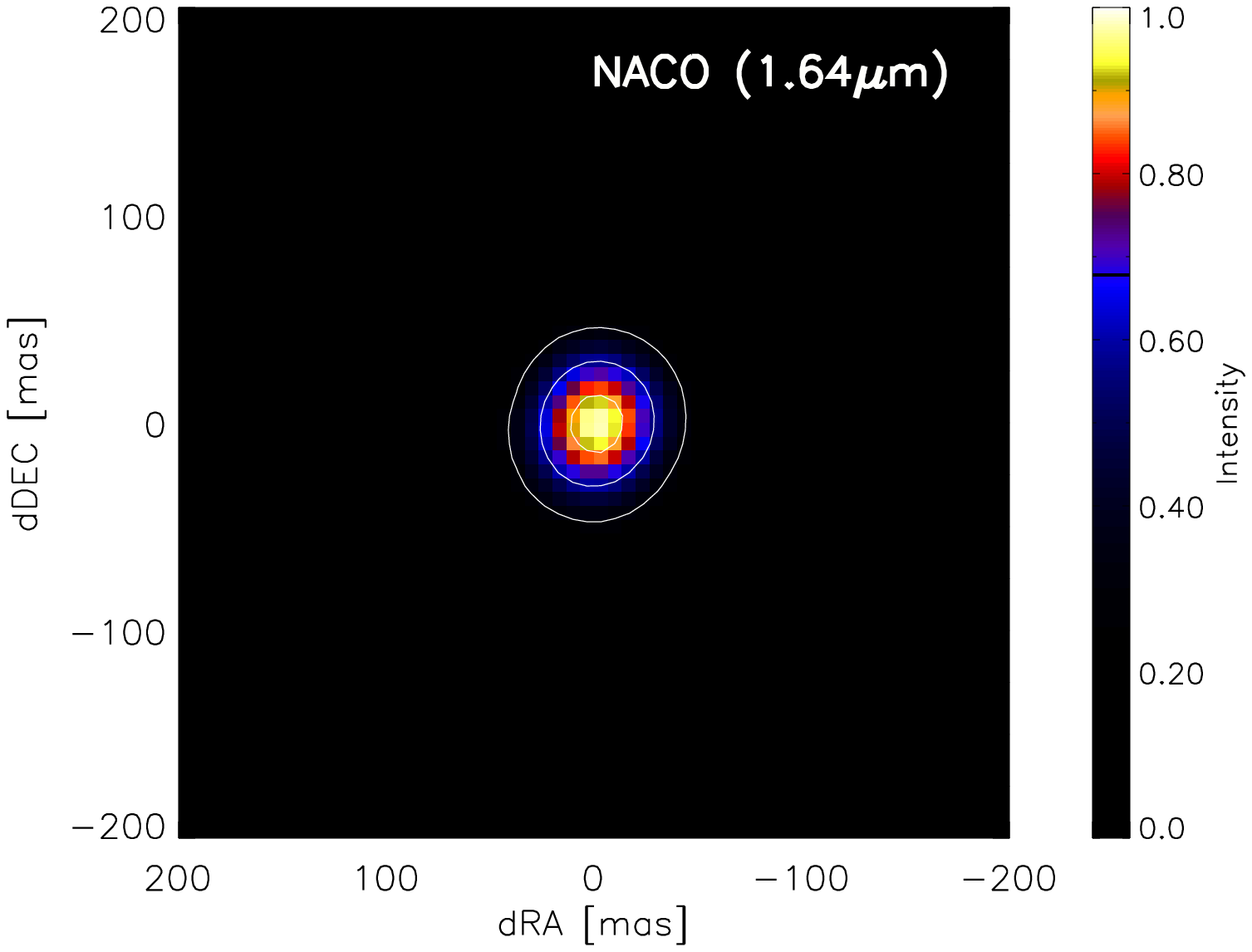} &
    \includegraphics[angle=0,scale=0.33, trim=0.4cm -1.3cm 3.1cm 0cm,clip=true]{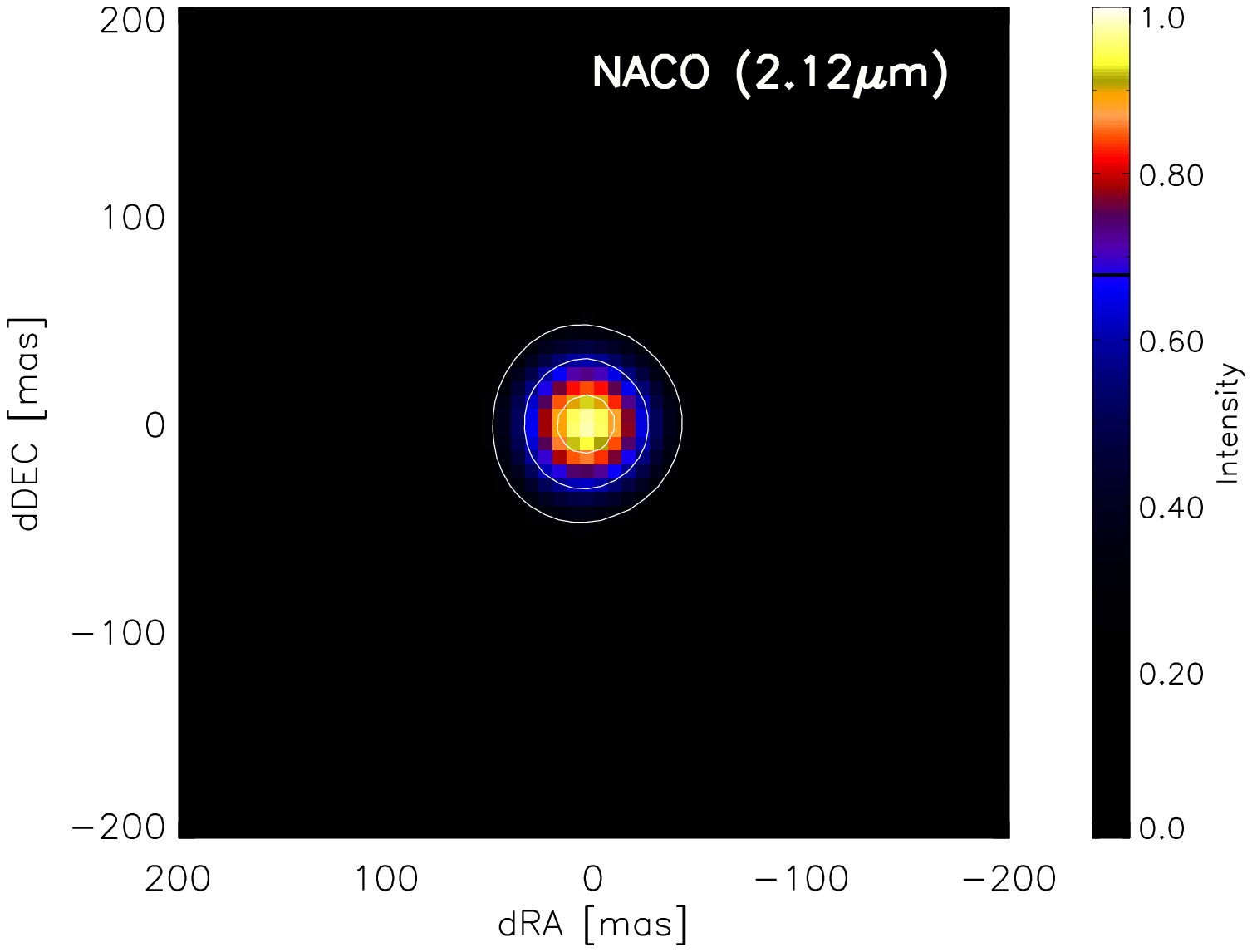} & 
     \multicolumn{2}{c}{\includegraphics[angle=0,scale=0.55]{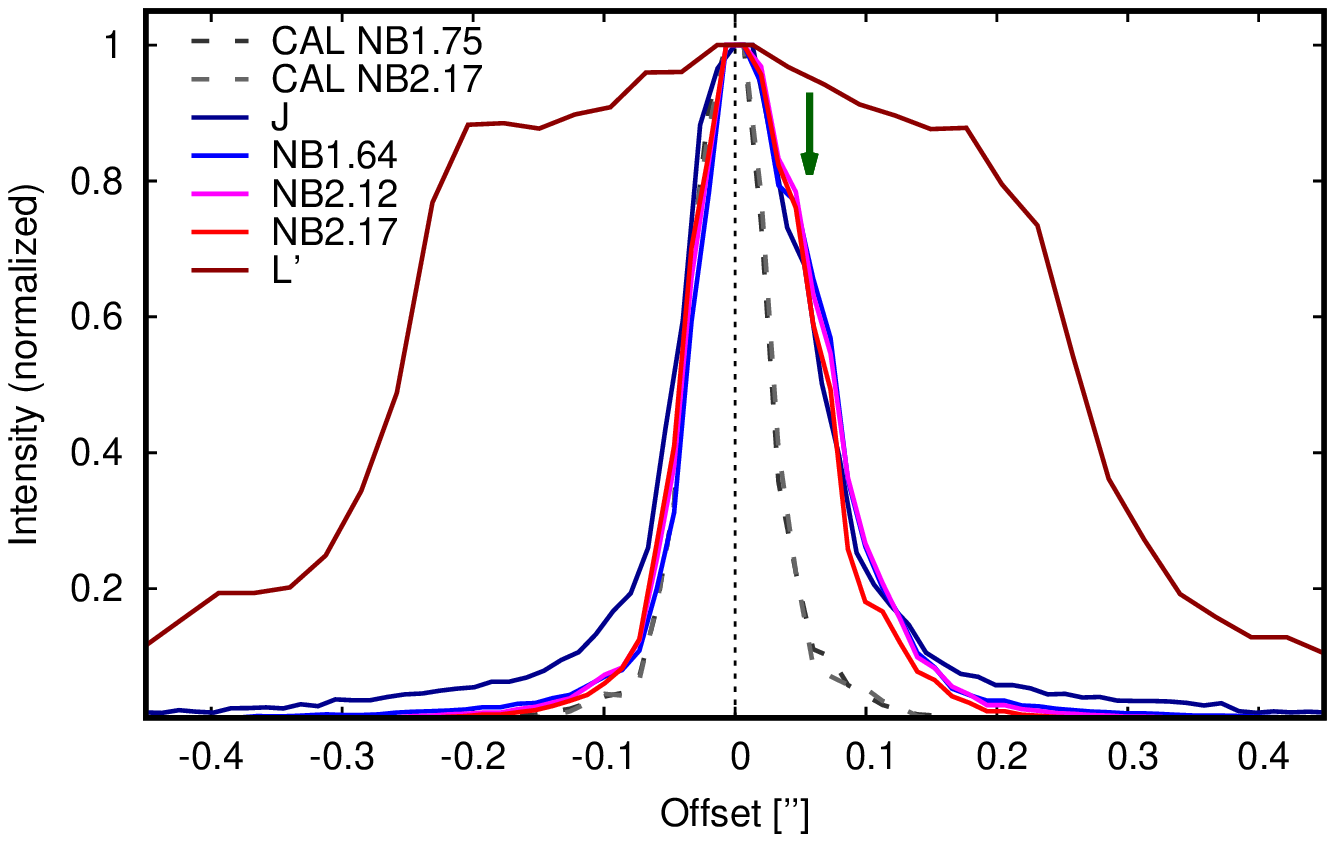}}
  \end{array}$
  \caption{
    NACO imaging of IRAS17216-3801 (top) and standard star HD155974 (bottom-left).
    For the J-band/1.64$\mu$m/2.12$\mu$m image, we plot intensity contours at 30\%, 60\%, and 90\% of peak intensity (white contours).
    The bottom right panel shows radial intensity cuts along the binary separation vector 
    from IRAS17216-3801 (solid lines) and the standard star images (dashed lines). 
    The arrow indicates the position of IRAS17216-3801-B, as derived from interferometry.\label{fig:NACO}}
\end{figure*}

We imaged the system on 2015-07-25 with the VLT/NACO adaptive optics system 
\citep{len03,rou03} 
using a $J$-band filter 
($8\times30$s integration) and a selection of narrowband filters 
centered on the {\FFeII} shock-tracing line 
(NB1.64; plus adjacent continuum with NB1.75 filter; $20\times1$s integration) 
and on the H$_2$ line (NB2.12; plus continuum with NB2.17 filter; $250\times1$s integration).  
For the narrowband filters, we interlayed the individual pointings with 
calibrator observations on HD155974 to determine the point spread function.
The system was also observed with a $L'$-band filter during our 2016-03-28 run
($1500\times0.2$s integration).  To eliminate the thermal background,
we dithered the object between the different detector quadrants.
The images were reduced using the NACO pipeline (Release~4.4.1).
The continuum-subtracted {\FFeII} and H$_2$ images do not exhibit
extended line emission and are therefore not further discussed.

\subsection{VLT/CRIRES spectro-astrometry}

We obtained VLT/CRIRES high-resolution spectroscopy \citep{kae04} near the Br$\gamma$~2.166$\mu$m line
on 2012-07-05. 
The long-slit spectra
offer spectral resolution $R=100,000$. 
Our setup was optimized for the spectro-astrometry technique,
where we measure the centroid position along the slit direction with high precision 
in order to derive the photocenter offset in the line with respect to the continuum.
We recorded spectra for six slit orientations (PA=0,60,120$^{\circ}$
plus anti-parallel orientations; each with $4\times45$s integrations), 
which allows us to remove spurious instrumental effects. 
Spectra were recorded on the calibrator Hip88154 to correct for telluric lines.
Details of our data processing procedure and on our
spectro-astrometric observable (X) can be found in \citet{kra12c}.

\begin{deluxetable*}{lcccc}
\tabletypesize{\scriptsize}
\tablecolumns{5}
\tablewidth{0pc}
\tablecaption{Model-fitting results (Sect.~\ref{sec:modeling})\label{tab:modelfitting}}
\tablehead{
                                & &  & \colhead{GAUSS model}       & \colhead{DISK model}  \\
}
\startdata
Binary separation                             & $\rho$       &  [mas]                                                 &  $57.94{\pm}0.24$         & $57.93{\pm}0.16$ \\ 
Binary PA                                           & $\Theta$ &  [$^{\circ}$]                                            &  $166.8{\pm}0.2$           & $166.76{\pm}0.2$ \\ 
PA change 2012-2016                    & $\Delta\Theta$ &  [$^{\circ}$]                                 &  $7{\pm}2$                      & $7{\pm}2$ \\
Flux ratio                                          & ${F_{\mathrm{A}}}/{F_{\mathrm{B}}}$    &                      &  $1.29{\pm}^{+0.09}_{-0.02}$     & $1.26{\pm}^{+0.13}_{-0.02}$ \\  
Extended flux contribution            & ${F_{\mathrm {ext}}}/{F_{\mathrm{tot}}}$  &                   &  $0.16{\pm}0.09$           & $0.16{\pm}0.06$\\
\hline
\multicolumn{5}{c}{\bf Circumprimary disk, continuum (Northern component, A)}\\
Contribution extended flux             & ${F_{\mathrm{A}}^{\mathrm{cs}}}/{F_{\mathrm{A}}}$ &  &  $0.60{\pm}0.08$          & $0.64{\pm}0.04$ \\
Disk inclination                                  & $i_{\mathrm{A}}$ &  [$^{\circ}$]                                 &  $89{\pm}10$                 & $60{\pm}10$ \\
Disk PA                                                & $\theta_{\mathrm{A}}$ &  [$^{\circ}$]                       &  $64{\pm}9$                   & $67{\pm}7$ \\
Gaussian FWHM size                         & $\Sigma_{\mathrm{A}}$ & [mas]                              &  $7.63{\pm}0.8$            & -- \\
Inner disk radius                                & $r_{\mathrm {A}}^{\mathrm{in}}$ &  [mas]                  &  --                                    & $2.77{\pm}0.39$ \\
Outer disk radius                               & $r_{\mathrm {A}}^{\mathrm{out}}$ &  [mas]                &  --                                    & $>12$   \\
\hline
\multicolumn{5}{c}{\bf Circumsecondary disk, continuum (Southern component, B)}\\
Contribution extended flux             & ${F_{\mathrm{B}}^{\mathrm{cs}}}/{F_{\mathrm{B}}}$ &  &  $0.90{\pm}0.05$           & $0.96{\pm}0.09$ \\  
Disk inclination                                 & $i_{\mathrm{B}}$ &  [$^{\circ}$]                                 &  $42{\pm}7$                    & $38{\pm}10$ \\
Disk PA                                               & $\theta_{\mathrm {B}}$ &  [$^{\circ}$]                       &  $185{\pm}32$               & $159{\pm}15$ \\
Gaussian FWHM size                         & $\Sigma_{\mathrm{B}}$ & [mas]                              &  $4.60{\pm}0.4$             & -- \\
Inner disk radius                                & $r_{\mathrm{B}}^{\mathrm{in}}$ &  [mas]                   &  --                & $2.49{\pm}0.42$\\
Outer disk radius                               & $r_{\mathrm {B}}^{\mathrm{out}}$ &  [mas]                 &  --               & $>10$ \\ 
\hline
                                                            & $\chi^{2}_{\mathrm{r}}$           &                               & 1.71           & 1.11
\enddata 
\tablecomments{
We define inclination $0^{\circ}$ as face-on orientation.  
PAs are measured East of North.}
\end{deluxetable*}

\section{Aperture synthesis imaging}
\label{sec:imaging}

To derive the brightness distribution from the VLTI data 
in a model-independent fashion we used the IRBis image reconstruction algorithm \citep{hof14}.
This algorithm minimizes a global cost function that includes a 
likelihood term (which determines the likelihood that the image is a representation of the data)
plus a regularisation term that prevents the algorithm 
from overfitting the data in the presence of an incomplete $uv$-coverage.

We first apply the algorithm to the GRAVITY data alone.
The resulting image (Fig.~\ref{fig:VLTI}, bottom-left) shows a binary with separation 
$\sim58$mas (170au at 3.1\,kpc) and a $K$-band contrast of $\sim0.6$.

We were also able to reconstruct an image from the combined 
2016/GRAVITY and 2012/AMBER data set, after accounting for the orbital motion 
between the two epochs.
For this purpose, we use the method outlined in \citet{kra05}, where
the $uv$-plane is rotated and scaled synchronously to the system motion.
In order to quantify the separation change $\Delta\rho$ and position angle (PA) change $\Delta\theta$
between 2012 and 2016, we varied these parameters systematically on a grid and 
searched for the image reconstructions with the lowest $\chi^2$.
We find a $\chi^2$-minimum with a anti-clockwise rotation of 
$\Delta\theta=8.5^{\circ}$
between the two epochs, while the separation stayed constant within the uncertainties.
This procedure results in the image shown in Fig.~\ref{fig:VLTI} (bottom-right) that represents the 
IRAS17216-3801 system at the 2016 epoch with a $3\times5$mas beam.

The binary is also detected in the 
NACO adaptive optics images obtained with the $J$-band, NB1.64, 
NB1.75, NB2.12, and NB2.17 filters, where we see some indications that the 
contrast increases towards shorter wavelengths, indicating that the northern
component has a redder color than the southern component (Fig.~\ref{fig:NACO}).

In all images, the northern component appears brighter than the southern component,
which leads us to denote the northern component with IRAS17216-3801-``A'' 
and the southern component with ``B''.
Comparing the intensity profile of the binary components in the image with
the interferometric beam clearly indicates that the emission is spatially extended,
both in the GRAVITY and GRAVITY+AMBER image (Fig.~\ref{fig:VLTI}).
The strong elongation seen in both components indicates that the circumstellar
disks are seen under inclination. 
The extended flux around A is elongated along 
PA$\sim70^{\circ}$, while B is elongated along PA$\sim0^{\circ}$.
Some artefacts are visible in the AMBER+GRAVITY images that result from the remaining imperfections
in the $uv$-coverage. These artefacts reach up to 10\% level of peak intensity,
but are easy to discern, as they mainly cause a repetition of the source binary pattern
along the south-east/north-west direction.

The emission from the circumstellar disks might also have been traced
by VLTI/MIDI interferometry \citep{bol13}.  
This data could be fitted with an over-resolved flux component (52\% of total flux) 
plus an elongated component with a Gaussian FWHM of $85^{+20}_{-33}$mas oriented along 
PA=$125^{+40}_{-10}$$^{\circ}$ \citep[inclination $59_{-11}^{+5}$$^{\circ}$,][]{bol13}.
Within the large uncertainties, the size and orientation are consistent with the 
derived binary separation vector, which leads us to suggest that the compact 
MIDI component traced IRAS17216-3801-A+B, but did not fully resolve the components.

\section{Modeling}
\label{sec:modeling}

\subsection{Spectral energy distribution}
\label{sec:sed}

To assist our further interpretation, we estimated the fundamental parameters 
of IRAS17216-3801 by compiling its spectral energy distribution (SED)
using archival data.  
Integrating over the SED yields a bolometric luminosity $L_{\mathrm{bol}}=6.1\times10^{4}$\,L$_{\sun}$
\citep[assuming d=$3.08$\,kpc;][]{bol13}.

We then fitted the SED with the radiative transfer model grid by \citet{rob06} and find a 
best-fit model (ID \#3009730) with an age of $\sim10^{5}$\,yrs, a $24M_{\sun}$ stellar mass
and a massive disk (gas+dust mass $1.2\times10^{-2}M_{\sun}$) 
seen at intermediate inclination ($50^{\circ}$).  
This SED model does not incorporate the companion, but we 
assume that it still provides a reasonable first-order estimate, 
as the primary will dominate the bolometric luminosity.
In order to estimate the mass of the individual components,
we assume that the measured $K$-band contrast is representative for the 
bolometric luminosity split between the two components and estimate from 
the mass-luminosity relation ($L{\propto}M^{3}$) the primary mass 
to $20M_{\sun}$ and the secondary mass to $18M_{\sun}$.
Spatially resolved multi-wavelength observations will be necessary to 
improve on this very rough estimate.

\subsection{Structure of circumstellar+circumbinary disks (continuum)}
\label{sec:modelingcont}

After identifying the basic morphology of the source with aperture synthesis imaging,
we fit geometric models to the visibilities/phases, with the goal to determine the 
precise binary astrometry and to characterize the circumstellar emission.

Our model includes emission from 
two point sources (separation~$\rho$; PA~$\Theta$),
where we allow the algorithm to associate each 
component with spatially extended flux. 
We denote the contributions of the spatially extended circumstellar emission to the
integrated flux of the northern and southern component with 
$F_{\mathrm{A}}^{\mathrm{cs}}/F_{\mathrm{A}}$ 
and $F_{\mathrm{B}}^{\mathrm{cs}}/F_{\mathrm{B}}$, respectively.
The flux ratio between the integrated fluxes 
is given by $F_{\mathrm{A}}/F_{\mathrm{B}}$.
In addition, we allow the algorithm to attribute a fraction of the total flux
($F_{\mathrm{ext}}/F_{\mathrm{tot}}$) to an over-resolved emission component.

We then fit the GRAVITY and AMBER data simultaneously
(using 50s temporal binning),
where we allow for orbital motion ($\Delta\Theta$) 
between the two epochs (2012.378 and 2016.476).
We explore the parameter space using an
Levenberg-Marquardt algorithm and vary the initial guess parameters
on a grid to avoid local minima.

For the brightness distribution of the circumstellar components, we assumed
either elongated Gaussians (``GAUSS'' model) or a temperature-power law disk model (``DISK'' model).
The GAUSS model introduces three free parameters to define the geometry 
of the circumstellar emission for each component, namely the 
inclination ($i_{\mathrm{A;B}}$), the PA ($\theta_{\mathrm{A;B}}$),
and the full-width-at-half-maximum ($\Sigma_{\mathrm{A;B}}$).  
The DISK model introduces four free parameters for each component, 
namely disk inclination, disk PA, 
and the inner and outer disk truncation radius 
($r_{\mathrm{A;B}}^{\mathrm{in}}$, 
$r_{\mathrm{A;B}}^{\mathrm{out}}$). 
The radial temperature gradient of the disk is parameterized as 
$T(r)=1500\mathrm{K}{\cdot}T(r_{\mathrm{in}})^{-0.43}$, 
where 1500\,K corresponds to the commonly assumed Silicate dust sublimation
temperature and $r^{-0.43}$ to the theoretically derived temperature gradient 
for irradiated flared dust disks \citep{chi97}.  Our DISK model implementation
has already been successfully applied to fit interferometry and SED data
of intermediate-mass and high-mass YSOs \citep{kra09b,kra10}.

The best-fit model (illustrated in Fig.~\ref{fig:data}, top-right) indicates that the disk 
around the southern component 
($\theta_{\mathrm{B}}=159\pm15^{\circ}$) is roughly aligned with the binary separation vector 
($\Theta=166.76^{\circ}$), while the northern component is oriented roughly 
perpendicular to it ($\theta_{\mathrm{A}}=67\pm7^{\circ}$).
For the radial intensity profile, we find that the DISK model provides 
a better representation of the data ($\chi^2_{\mathrm{r}}=1.11$)
than the GAUSS model ($\chi^2_{\mathrm{r}}=1.71$), 
indicating that the disks are not smooth, but exhibit a central brightness depression, 
potentially tracing the opacity drop associated with dust sublimation near the star.  
The measured inner disk truncation radii of $2.78$ and $2.49$mas (8.6 and 7.7au) 
are consistent with the theoretical dust sublimation radii, if one assumes grey dust
and dust sublimation temperature of 1300K.

Comparing the properties of component~A and B, we find that component~A 
contributes a larger fraction to the total $K$-band flux, has a larger inner disk cavity, 
and a higher point-source flux contribution, all pointing to the conclusion that 
this is the more luminous and more massive component in the system.

Our best-fit model attributes 66\% of the total flux to the circumstellar disks,
$17\pm4$\% and $1\pm6$\% to photospheric/unresolved emission close to the stars,
while $16\pm6$\% extended emission that is over-resolved on the shortest baselines,
indicating a Gaussian half-width-at-half-maximum (HWHM) size $\gtrsim50$mas.
We speculate that this extended $K$-band flux might trace the
same physical structure that is also seen in our $L'$-band image.
The radial intensity cut through the $L'$-band image 
reveals that IRAS17216-3801-A+B contribute $\lesssim20$\%, 
while the majority of the $L'$ flux traces an 
extended component with a HWHM size $\sim200$mas (Fig.~\ref{fig:NACO}, bottom-right). 
We note that this measured size is in good agreement with the expected inner
truncation radius of the circumbinary disk around IRAS17216-3801-A+B.
Hydrodynamic simulation suggest that binary systems with low eccentricity
truncate their circumbinary disk at $\sim2-3\times$ the semi-major axis $a$.
Using our lower limit on the semi-major axis ($a\gtrsim58$mas),
we therefore expect the circumbinary disk to extend beyond $\gtrsim 120$-180mas,
in good agreement with the size of the $L'$-band structure.

\subsection{Accretion-tracing and disk-tracing emission lines}
\label{sec:modelingline}

\begin{figure*}[p]
  \centering
     $\begin{array}{c@{\hspace{2mm}}c}
        \parbox[c]{6.4cm}{
        \includegraphics[angle=0,width=6.3cm]{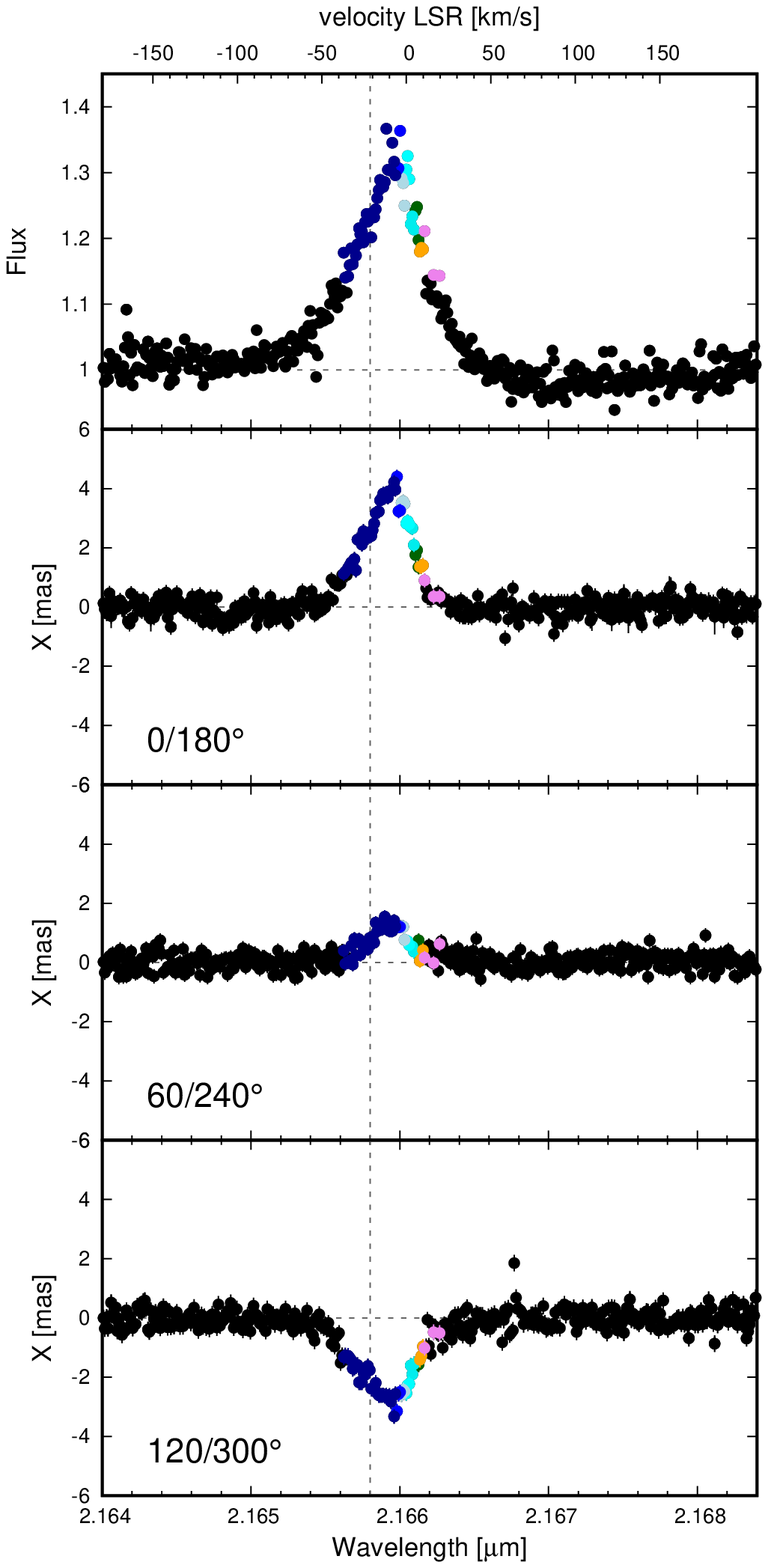} \newline
        \includegraphics[angle=0,width=6.4cm]{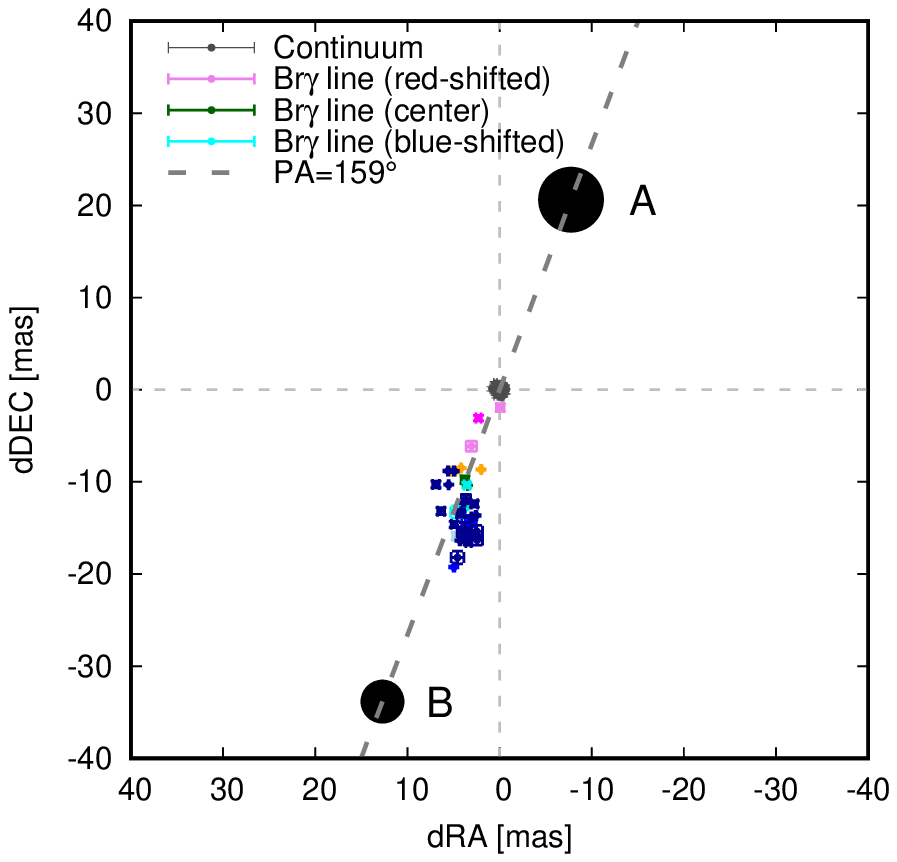}}
        &
          \parbox[c]{11.5cm}{
          \includegraphics[angle=0,width=11.3cm]{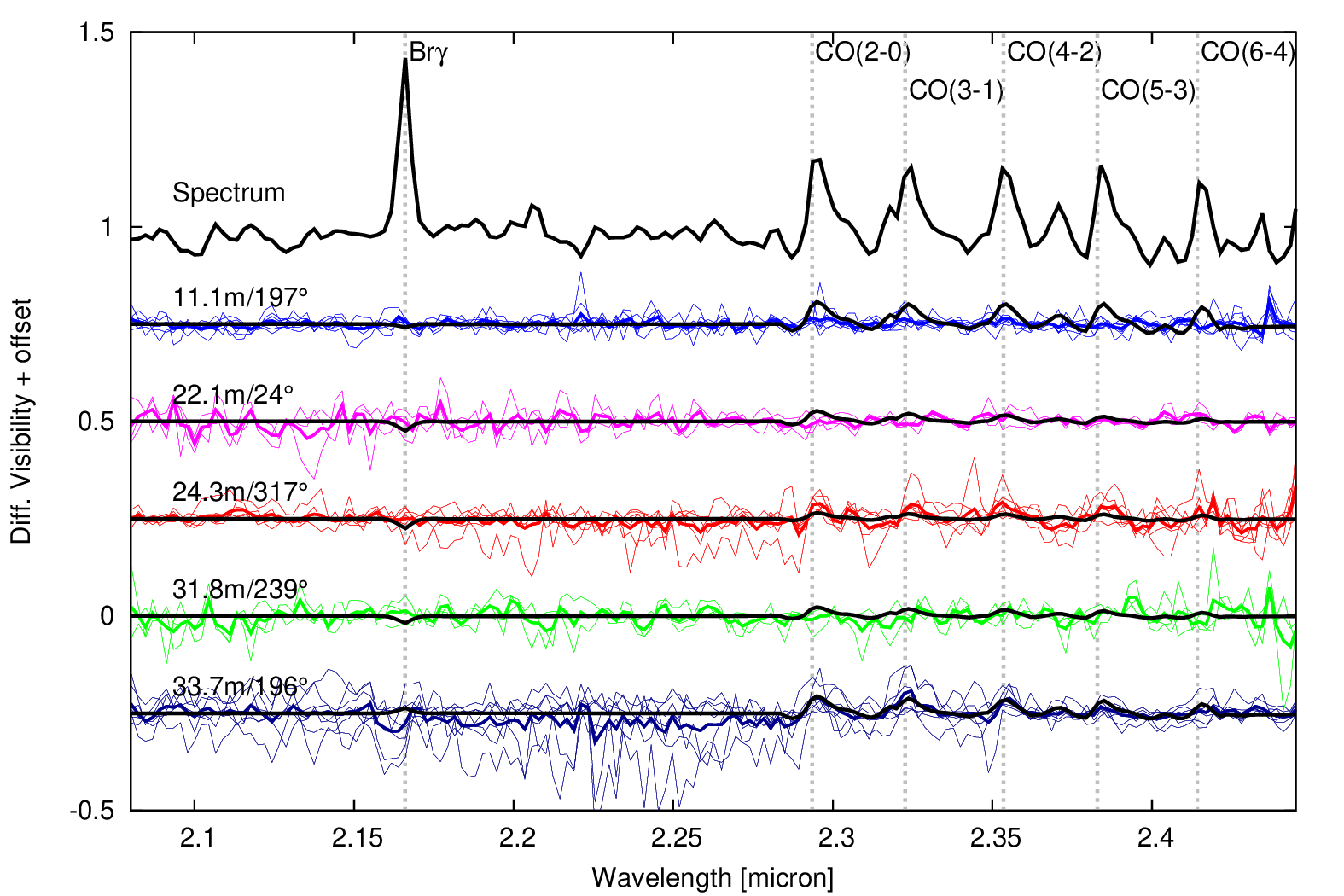} \newline
          \includegraphics[angle=0,width=11.3cm]{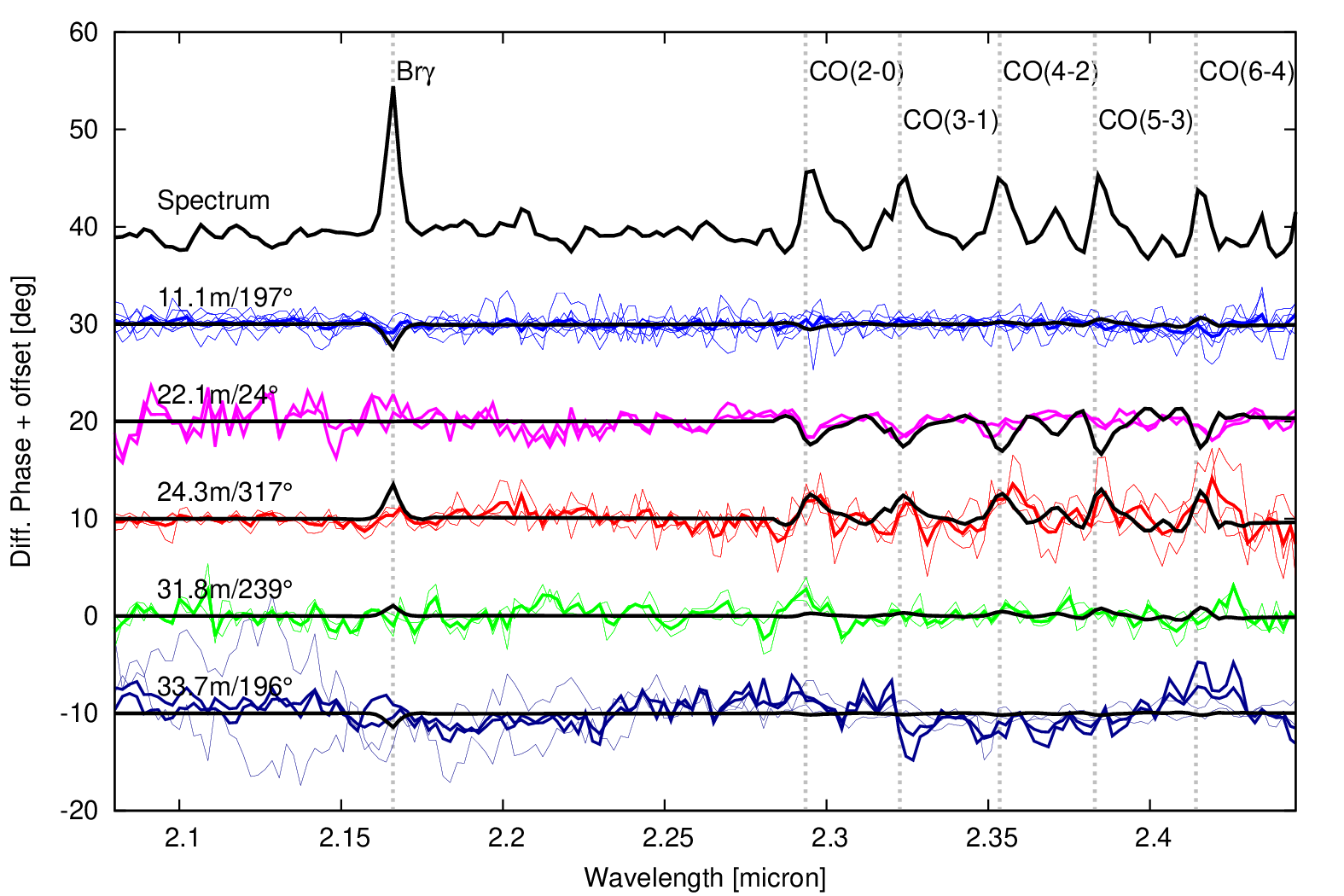} 
          }
      \end{array}$
  \caption{
    CRIRES spectro-astrometric (left) and GRAVITY spectro-interferometric data (right) on IRAS17216-3801.
    Top-left: The Br$\gamma$ line is single-peaked and shows strong astrometric signals (X) toward different PAs.
    Bottom-left: Continuum-corrected 2-dimensional photocenter offsets derived from the astrometric signals, with the color
    indicating the wavelength of the corresponding channel.
    For clarity, we also indicated the position of IRAS17216-3801-A+B, as derived with GRAVITY+AMBER (Sect.~\ref{sec:modelingcont}).
    Right: Spectra, differential phases, and differential visibilities derived from our GRAVITY data.
    The black lines show the visibilities/phases that correspond to the Br$\gamma$/CO model outlined in Sect.~\ref{sec:modelingline}.\label{fig:lines}}
\end{figure*}

The GRAVITY wavelength-different visibilities/phases
allow us to determine the distribution of the Br$\gamma$ and CO line-emitting gas.
It is interesting to note that the visibility/phase signals in Br$\gamma$ and CO 
appear to be uncorrelated and are stronger in CO (despite their weaker line strength),
which already indicates that these lines emerge from substantially different regions 
in the circumstellar environment. 

The Br$\gamma$-line traces hot ($\gtrsim10^4$K) ionized gas 
and is a well-established mass accretion/ejection tracer
\citep{nat04,kra12c,car16}.
Therefore, we model Br$\gamma$ as unresolved emission at the 
position of the stars.  Our modeling allows us to exclude scenarios that associate 
the Br$\gamma$ only with the primary or secondary component, 
as this results in much stronger visibility/phase signals than measured.
We achieve a satisfactory fit by attributing $40\pm10$\% of the Br$\gamma$-emission
to the primary and $60\pm10$\% to the secondary (Fig.~\ref{fig:lines}, right).

The CRIRES spectro-astrometry provides an independent method for
constraining the origin of the Br$\gamma$ emission.
The Br$\gamma$ photocenter is displaced along PA=$161\pm3^{\circ}$, 
which is consistent with the binary separation vector 
derived from the AMBER data for the same epoch ($159.4^{\circ}$).
The offset is $\sim14\pm1$mas, which places the photocenter
between the two stars (Fig.~\ref{fig:lines}, left),
again indicating that both components contribute to the line emission.
Modeling the photocenter displacement quantitatively we find that the 
primary (secondary) contributes 38\% (62\%) of the Br$\gamma$-flux.
Therefore, the GRAVITY and CRIRES data indicate independently 
that the (likely lower-mass) B component accretes at a higher rate.

The CO bandhead emission is believed to trace warm ($\gtrsim10^3$K) 
neutral gas in the more extended disk regions
\citep[few au to a few hundred au in massive YSOs;][]{ile13}.
To model the GRAVITY CO line data, we approximate the 
CO-emitting region with a Gaussian of FWHM $\Sigma_{\mathrm{CO}}$.
We vary the position of the Gaussian along the binary separation vector,
where $s_{\mathrm{CO}}$ denotes the separation of the Gaussian to the primary.
We find that the CO emission is extended ($\Sigma_{\mathrm{CO}}=5.8\pm0.8$mas)
and that it is centered on a position between the two stars ($s_{\mathrm{CO}}=31.6\pm3.3$mas),
possibly including contributions from both circumstellar disks and the
gas streams between the disks (Fig.~\ref{fig:lines}, right).

\section{Conclusions}
\label{sec:conclusions}

Our VLTI interferometry resolves the high-mass protostar 
IRAS17216-3801 into a close (58mas=170au) binary.
The system is $\sim3\times$ as massive ($\sim20+18M_{\sun}$) 
and $\sim5\times$ more compact than the benchmark protobinary IRAS20126+4104 
\citep[separation {850au,}][]{sri05}.
IRAS17216-3801 is also $\sim3\times$ more compact than the 
NGC7538~IRS1a/b system (separation {500au}), where 
Keplerian-rotating methanol maser disks \citep{pes04,mos14} 
and ammonia line absorption consistent with a circumbinary envelope were found \citep{god15}.

Our IRAS17216-3801 imaging traces the thermal dust emission in
the circumprimary and circumsecondary disk and reveals that both disks
are misaligned with respect to the binary separation vector. 
Various formation mechanisms have been proposed that might have
resulted in the observed disk misalignments, 
including turbulent disk fragmentation, perturbation 
by a third component, star-disk capture, or infall of material whose 
angular momentum vector was misaligned to that of the gas 
that formed the binary initially \citep{bat10}.  
Once a highly misaligned system has formed, tidal interactions 
will work towards re-aligning the disks, which has
been proposed to happen on the viscous timescale \citep{pap95} or 
on the much shorter precession timescale \citep{bat00}.
From the deduced stellar masses and outer disk radii (Tab.~\ref{tab:modelfitting})
we estimate the precession timescale of the IRAS17216-3801 circumprimary disk 
to $\sim200,000$~years and $\sim900,000$~years for the circumsecondary disk,
although these values should be considered upper limits as we derive only 
lower limits on the disk radii from our infrared observations.
Therefore, our observations indicate that the tidal realignment is 
still ongoing, highlighting the young dynamical age of the system.

The stronger orbital misalignment of the circumprimary disk 
might also explain its larger spatial extend, as the misalignment results
in a weaker Lindblad torque that acts to truncate the circumstellar disks 
\citep{lub15}.

The circumstellar disks are fed by a circumbinary disk/envelope
that has been resolved by our VLT/NACO $L'$-band imaging.
We detect an extended structure, whose size matches roughly the 
expected dynamical truncation radius of the IRAS17216-3801 binary system.
Both circumstellar disks show signatures of ongoing accretion,
where we measure a higher Br$\gamma$-line luminosity 
at the position of the secondary than at the primary.
This suggests that the secondary disrupts the accretion stream
on the primary and channels most of the infalling material
onto the circumsecondary disk, confirming the prediction of
hydrodynamic simulations \citep{bat97}.
We resolved for the first time the CO line-emitting region in a massive YSO
and find that it traces primarily warm neutral gas located between the two components.

With its unique properties
IRAS17216-3801 provides an ideal laboratory for studying the 
formation of young high-mass multiple systems and to 
unravel how such systems accrete from their disks.

\acknowledgments

We thank the GRAVITY consortium and the Science Verification team, which is composed of 
ESO employees and GRAVITY consortium members (\url{https://www.eso.org/sci/activities/vltsv/gravitysv.html}).
We acknowledge support from an STFC Rutherford fellowship/grant (ST/J004030/1, ST/K003445/1),
Marie Sklodowska-Curie CIG grant (\#618910), 
Philip Leverhulme prize (PLP-2013-110), and
ERC Starting grant (Grant Agreement \#639889).

{\it Facilities:} \facility{VLTI}, \facility{VLT}.


\end{document}